# Precise tuning of the superconducting properties of Mn-doped Al films for transition edge sensors by ion-implantation

Yue Lv[1,2,4]\*, Hao Huang[1,2,4,5]\*, Tiangui You[1,4], Feng Ren[3], Xin Ou[1,4]†, Bo Gao[1,2,4]†, Zhen Wang[1,2,4]

1. State Key Laboratory of Functional Materials for Informatics, Shanghai Institute of Microsystem and Information Technology, Chinese Academy of Sciences, 865 Changning Road, Shanghai 200050, China.

2. CAS Center for Excellence in Superconducting Electronics (CENSE), Shanghai 200050, China

3. School of Physics and Technology, Wuhan University, Wuhan 430072, China

4. Center of Materials Science and Optoelectronics Engineering, University of Chinese Academy of Sciences, Beijing 100049, China

5. School of Physical Science and Technology, ShanghaiTech University, 201210, Shanghai, China

\*These authors contribute equally to this work.

†Email: ouxin@mail.sim.ac.cn, bo_f_gao@mail.sim.ac.cn

**Abstract**

**Magnetic impurities in metallic superconductors are important for both fundamental and applied sciences. In this study, we focused on dilute Mn-doped aluminum (AlMn) films, which are common superconducting materials used to make transition edge sensors (TES). We developed a multi-energy ion-implantation technique to make AlMn films. Compared with frequently used sputtering techniques, ion-implantation provides more precise and reliable control of the Mn doping concentration in the AlMn films. The ion implantation also enables us to quantitatively analyze the superconducting transition temperature curve as a function of the Mn doping concentration. We found that Mn dopants act as magnetic impurities and suppression of superconductivity is counteracted by the antiferromagnetic Ruderman–Kittel–Kasuya–Yosida interaction among Mn dopants. The RKKY interaction can be tuned through defect engineering in the ion-implantation process and through post-implantation annealing.**

# I. INTRODUCTION

The problem of magnetic impurities in metallic superconductor hosts has been of great interest for a long time[1-5]. Understanding interactions among magnetic impurities in such hosts poses important questions at the interface of fundamental and applied sciences[6-20]. Magnetic impurities are generally thought to suppress superconductivity; however, the actual pair breaking process can be dominated by different mechanisms. A quantitative treatment of this problem is based on the Anderson model of impurity states[1], which considers a localized transition metal dopant with a d-like orbital carrying a magnetic moment embedded in a free electron gas. The hybridization of the impurity state with the conduction-electron states of the host metal leads to broadening of the d-like density of states (DOS). Abrikosov and Gorkov (AG) studied this model in the weak-scattering approximation[2], in which the broadening of the d-like DOS is much smaller than the energy difference between spin-up and -down states owing to Coulomb repulsion. They found that impurities remain magnetic. The spins of magnetic impurities interact with one another via their intrinsic dipolar interaction and the effective interactions among impurities are enhanced by the exchange coupling of impurity spin to conduction electrons. For a simple metal host, this results in a Ruderman–Kittel–Kasuya–Yosida (RKKY) coupling among magnetic impurities. The suppression of the superconducting transition temperature ($T_c$) of the metallic superconductor host can be strengthened or counteracted by the RKKY coupling, depending on whether it is ferromagnetic or antiferromagnetic[6]. Eventually the suppression of $T_c$ leads to gapless superconductivity. Conversely, transition metal dopants can also behave as non-magnetic impurities if the scattering rate of conduction electrons is sufficiently high that the broadening of the d-like DOS exceeds the energy difference between the spin-up and -down states. In these circumstances, the magnetic moments of the transition metal dopants are washed out by the conduction electrons of the metal host. Kaiser et al. addressed this case and found that non-magnetic impurities destroy superconductivity of the host by suppressing electron-phonon coupling strength through the pair-breaking effect of Coulomb repulsion between opposite spin d electrons, without breaking the time-reversal symmetry of Copper pairs[4]. Therefore, the host superconductor retains its Bardeen–Cooper–Schrieffer (BCS)-like DOS as $T_c$ is suppressed.

In this letter, we focus on dilute Mn-doped Al (AlMn) films. We are particularly interested in AlMn because it is a common material for making transition edge sensors (TESs)[21-27], which are highly sensitive equilibrium photon detectors. Because of the excellent noise performance, TESs have an unreplaceable role in many cutting-edge scientific instruments, including ground-based cosmic microwave background (CMB) observatories dedicated to the search for the primordial gravitational waves, which remain the most important untested hypothesis of inflation. Magnetically doped superconducting films are excellent candidates to make TESs and Mn-doped aluminum is a representative material of this kind. Currently, magnetron-sputtering methods are mainly used to grow AlMn films[10,28,29]. Because it is difficult to precisely control the Mn concentration in the sputtering target and inconvenient to frequently replace the target, tuning the $T_c$ of AlMn films relies on post-deposition annealing[24]. To find a more precise and reliable way to control $T_c$ and superconducting homogeneity of the AlMn films, we used an ion-implantation, which is a mature technique of the semiconductor industry that has been widely used to fabricate semiconducting image sensors, as reviewed by Teranishi et al[30]. Recently, this technique has also been used to improve the performance of superconducting nano-strip detectors through defect engineering induced by He-ion implantation[31]. In the past, several groups have used ion-implantation to introduce magnetic dopants, such as Fe and Mn, into W, Al, and Ti films[32-36]. In these studies, suppression of the transition temperatures was observed; however, a need remains for thorough investigation of the electric properties of doped superconducting films with respect to various process parameters, such as the energy and the fluence of the implantation process and the film thickness. Notably, a previous study on sputtered AlMn films showed that Mn dopants are non-magnetic based on the observation of a sharp BCS-like DOS in tunneling experiments[11]. Because the magnetic moments of Mn dopants in an Al host are very sensitive to local lattice distortion[8], we wanted to investigate if Mn dopants behaved differently in these ion-implanted films, which may have an important influence on the superconducting properties of AlMn films.

We used a multi-energy implantation technique to make AlMn films, which is necessary to form a homogeneous distribution of Mn dopants in the vertical direction of the films. With this method, we demonstrate precise control of the Mn doping concentration and the superconducting transition temperature of AlMn films. The transition temperature was reduced to be less than 100 mK, suggesting potential for the use of ion-implanted AlMn films to make TESs. Quantitative analysis of the superconducting transition temperature curve as a function of Mn doping concentration showed that Mn dopants act as magnetic impurities, and exchange coupling among these impurities is strongly influenced by the defects induced by ion-implantation.

## II. Preparation of Mn-doped Al films

The aluminum film was grown by magnetron sputtering on 4'' $SiO_2$/Si wafer. The substrate was cleaned with an argon ion beam before the Al deposition. The typical growth conditions were a DC sputtering mode at a constant current of 0.31 A, and a power of 115 W with a variation of 2 W, and pressure of 0.5 Pa. After growth of the aluminum film, the wafer was diced into 30-mm-large squares to fit the sample holders of the ion-implanters, and then the Mn ion implantation was performed, either at the Helmholtz-Zentrum Dresden-Rossendorf, Germany or in Wuhan University, China. Because single energy implantation generated a Gaussian distribution of Mn ions in Al films in the vertical direction, we used a multi-energy implantation technique to achieve a homogenous distribution of Mn ions in both lateral and vertical directions. For convenience, we selected an implantation plan based on three different implantation energies. The implantation energies were selected such that the Gaussian peaks corresponding to each energy coincided with the quadrisection points of the film thickness. For 100-nm-thick Al films, typical implantation energies were 30, 65, and 100 keV; for thicker films, higher implantation energies were needed so that Mn-ions could reach the interior of the Al film. Typical implantation energies for the 300-nm-thick film were 100, 200, and 300 keV. The implantation fluence for each energy was selected such that the variation of the simulated Mn concentration in the vertical direction with respect to an ideal homogenous distribution of Mn ions in the film was minimized. Figure 1 shows the simulated and measured Mn distributions in 300-nm-thick aluminum films. The target concentration of Mn dopant atoms was 1100 ppm. The three Gaussian-type dotted curves, with their peak position shifting from left to right, are the simulated Mn distributions corresponding to 100-, 200-, and 300-keV implantation energy determined by the simulation software SRIM-2008[37], respectively. The black solid curve is the sum of the three Gaussian curves and the red solid curve is the actual Mn distribution measured in the films by secondary-ion mass spectroscopy, which confirmed a homogeneous distribution of Mn ions in the vertical direction and the actual Mn doping concentration meets our expectation.

## III. Evolution of $T_c$ with Mn doping concentration

After the ion-implantation, the AlMn films were patterned into 660 μm long and 50 μm-wide slices for electrical characterization by photo-lithography and chemical wet-etching. The electrical measurements were performed in an adiabatic demagnetization refrigerator using a lakeshore AC resistance bridge (372 series) in a four-terminal measurement geometry. The current stimulus used in these measurements was 10 μA. Figure 2 shows the resistivity versus temperature curves of undoped and Mn-doped aluminum films. The thicknesses of these samples were 100, 150, and 300 nm, respectively. The superconducting temperatures of the undoped Al films, are approximately 1.2 K for all three different film thickness. The resistivity ratio ($\rho$(293K)/$\rho$(4.2K)) for the undoped Al films was above 5 for all cases, and the ratio for a 300-nm thick Al film even reached 7.6, suggesting the formation of high-quality films. All the Mn-doped samples had transition temperatures lower than 100 mK and retained a sharp resistive transition, with a maximum width less than 20 mK. These features are comparable with those of sputtered AlMn films[11]. The very low $T_c$ and sharp transition edge suggest potential for use of these films to make transition edge sensors.

To investigate how the superconducting transition temperature changes with the Mn doping concentration, we plot the dependence of $T_c$ on the Mn doping concentration in Figure 3. We focused mainly on the 100- and 300- nm thick films. The Mn doping concentration was not the only parameter that influenced the transition temperature of the AlMn films. The 300-nm-thick films were more sensitive to Mn-ion doping and the transition temperature decreased more rapidly as the Mn concentration increased. To better understand how Mn dopants affect the superconducting transition temperature of AlMn films, we quantitatively analyzed our data based on fitting the curves shown in Figure 3 with the use of the AG and Kaiser models. We used a general form of the transition temperature of superconductors containing non-magnetic and magnetic impurities[11]:

$$\beta + \ln(\frac{T_c}{T_{c0}}) + \psi(0.5 + \frac{\alpha}{2\pi k_B T_c}) - \psi(0.5) = 0, \qquad (1)$$

when the impurities had magnetic moments, as is the case for the AG model, $\beta = 0$ and the transition temperature were controlled by the pair-breaking parameter $\alpha$, which is

defined as:

$$\alpha = 0.882 k_B T_{c0} \frac{x}{x_c}[1-\delta \frac{x}{x_c}\frac{T_{c0}}{T_c}], \qquad (2)$$

where $\delta$ is the magnetic coupling parameter among magnetic impurities ($\delta < 0$ for ferromagnetic coupling and $\delta > 0$ for antiferromagnetic coupling), $x_c$ is the critical dopant concentration for which $T_c = 0$ (at $\delta = \beta = 0$), and $T_{c0}$ is the transition temperature of the undoped superconductor.

If the impurities are considered to be non-magnetic, as in the case of Kaiser model, we have $\alpha = 0$ and equation (1) reduces to

$$\beta + \ln(\frac{T_c}{T_{c0}}) = 0, \beta = \frac{1}{\lambda_{eff}}\frac{x/x_c}{1-x/x_c}, \qquad (3)$$

where $\lambda_{eff}$ is an effective electron-coupling (e-ph) strength parameter and $x_c$ is the critical dopant concentration for which $T_c = 0$ (at $\alpha = 0$). Table 1 lists the fitting parameters:

Table 1: Fitting parameters of the AG and Kaiser models

| *Film thickness* | | *100 nm* | *300 nm* |
|---|---|---|---|
| *Kaiser Model* | $\lambda_{eff}$ | 0.597 | 1.427 |
| | $x_c$ (ppm) | 4179 | 1680 |
| *AG Model* | $\delta$ | 0.0137 | 0.0033 |
| | $x_c$ (ppm) | 1956 | 1253 |

The value of $\lambda_{eff}$ for both 100- and 300-nm films, found in the fitting based on the Kaiser model, are greater than the value 0.15 found in previous studies of sputtered AlMn films[11], and also larger than the value 0.169 for pure aluminum[11]. These effective e-ph coupling parameters seem to be unrealistic because in the Kaiser model, we expect that the Coulomb repulsion will suppress the electron-phonon coupling strength and the value of $\lambda_{eff}$ should be reduced[4]. The fitting results of the AG model show that the antiferromagnetic coupling parameter $\delta$ of 300-nm thick AlMn films is lower than that of 100-nm thick films. The antiferromagnetic coupling counteracts the suppression of superconductivity by magnetic impurities[6], which explains why the $T_c$ of the 300-nm thick films decreased faster as the Mn doping concentration increased. Thus, we conclude that the Kaiser model might not be appropriate for our samples and that the Mn dopants (at least part of these dopants) in the ion-implanted AlMn films act as magnetic impurities.

## IV. Simulation of defects

The major difference in the fabrication processes of the 100- and 300-nm thick films was the implantation energy. Much higher implantation energies were used to fabricate the 300-nm thick AlMn films. We speculate that more defects were introduced in the higher energy implantation process, and those defects suppressed the RKKY coupling between Mn impurities. Ion implantation processes can damage host materials, introducing defects, such as interstitial atoms, substitution, and vacancies. Because it is difficult to specify which type of defects are responsible for the suppression of the RKKY coupling, we used the displacements per atom (DPA) to measure the damage to the host lattice by $Mn^+$-ion implantation. DPA is a dimensionless number defined as the average number of displacements of each host lattice atoms by irradiation ions and subsequent cascading collision. It is widely used to estimate radiation damage in materials. For a simple metal host, the total amount of DPA is determined with the use of the following formula[38]:

$$DPA = \int_0^{t_e} dt \int_0^\infty dE \sigma_d(E) \phi(E,t), \tag{4}$$

where $\sigma_d(E)$ is the atomic displacement cross-section, $\phi(E,t)$ is the radiation flux spectrum, and $t_e$ is the exposure time. We simulated the DPA using TRIM-2008 along the depth direction of two AlMn films; one 100-nm thick and the other 300-nm thick. The Mn doping concentration in both films was 1100 ppm. As shown in Figure 4, the dotted and dashed lines are the simulation curves corresponding to each individual implantation energy used in implantation of the 100- and 300- nm thick films, and the solid curves are the sum of the dotted/dashed curves. The radiation damage is clearly more severe for the 300-nm thick AlMn film.

## V. Effects of Annealing

To further demonstrate that defects induced by the ion-implantation influenced the RKKY coupling among the Mn dopants, we annealed the 300-nm thick AlMn films at 220 and 230 °C and then measured the superconducting transition temperatures. Figure 5 shows the transition temperature curves of these annealed samples, together with that from the unannealed samples. We also fit the curves of $T_c$ versus Mn concentration based on the AG model. As shown in the figure, the magnetic coupling parameter $\delta$ increases from 0.0033 (unannealed) to 0.0261 (220 °C) and 0.0417 (230 °C). We attribute the increase of the magnetic coupling parameter to healing of the defects through annealing. The post-implantation annealing reduced defects in the AlMn films, resuming the exchange coupling among Mn impurities and increasing the superconducting temperature of the AlMn films.

## VI. CONCLUSIONS

Our results suggest that Mn dopants atoms in ion-implanted Al films act as magnetic impurities. This result is different from previous work of O'Neil et al.[11], in which Mn doping into an Al lattice was realized by sputtering a Mn-doped Al target, and the authors concluded that Mn dopants are non-magnetic through observations of a sharp BCS-like density of states in tunneling experiments. Our findings show that the method of preparing AlMn films can strongly influence the interactions between the Mn dopants and Al host. An early study of ion-implanted Al films also suggested that Mn dopants in an Al matrix form a spin fluctuation system[36]. The applicability of the AG and Kaiser models thus depends on the AlMn film growth conditions. In conclusion, the ion-implantation technique enables us to precisely control the Mn doping concentration in AlMn films and to arbitrarily tune the superconducting transition temperature. AlMn films with a sharp transition edge and low resistivity are obtained, implying potential applications to making transition edge sensors. We find that Mn dopants act as magnetic impurities and the exchange coupling interactions among these impurities are sensitive to defects created in the ion-implantation, which are influenced by the film thickness, implantation energy, and post-implantation annealing. The antiferromagnetic RKKY coupling counteracts the suppression of the superconducting transition temperature by Mn dopants. Our results show that ion-implantation is a powerful technique in fundamental and applied research of dilute magnetically doped superconductors.

## Acknowledgements

This work is supported by National Key Research and Development Program of China under grant No. 2017YFA0304000, by Chinese National Science Foundation under grant No. 11653004, No. 11705262, No. 61874128, No. 61851406, No. U1732268 and No. U1632272, by Frontier Science Key Program of CAS (No.: QYZDY-SSW-JSC032), and by Program of Shanghai Academic Research Leader (No.: 19XD1404600). The nanofabrication work is supported by the superconducting electronics facility (SELF) of Shanghai institute of microsystem and information technology. We thank Wentao Wu, Hubing Wang and Qi Jia for their assistance in the experiments. We thank Helmholtz-Zentrum Dresden-Rossendorf (HZDR) for the implantation work.

# Figure 1

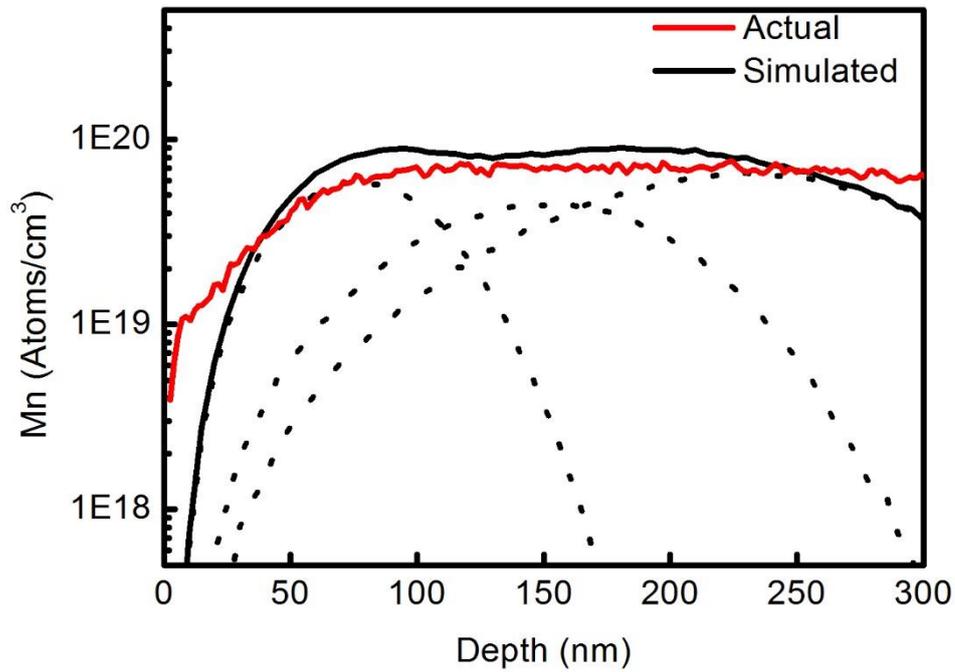

**Figure 1.** Simulation of the Mn doping concentration in the vertical direction of a 300-nm thick AlMn films. The three Gaussian type dotted curves are simulated Mn distribution curves corresponding to each implantation energy (100, 200 and 300 keV, from left to right, respectively). The black solid curve is the sum of the three Gaussian curves and the red solid curve is the Mn doping concentration measured by secondary-ion mass spectroscopy. The target Mn centration was 1100 ppm.

# Figure 2

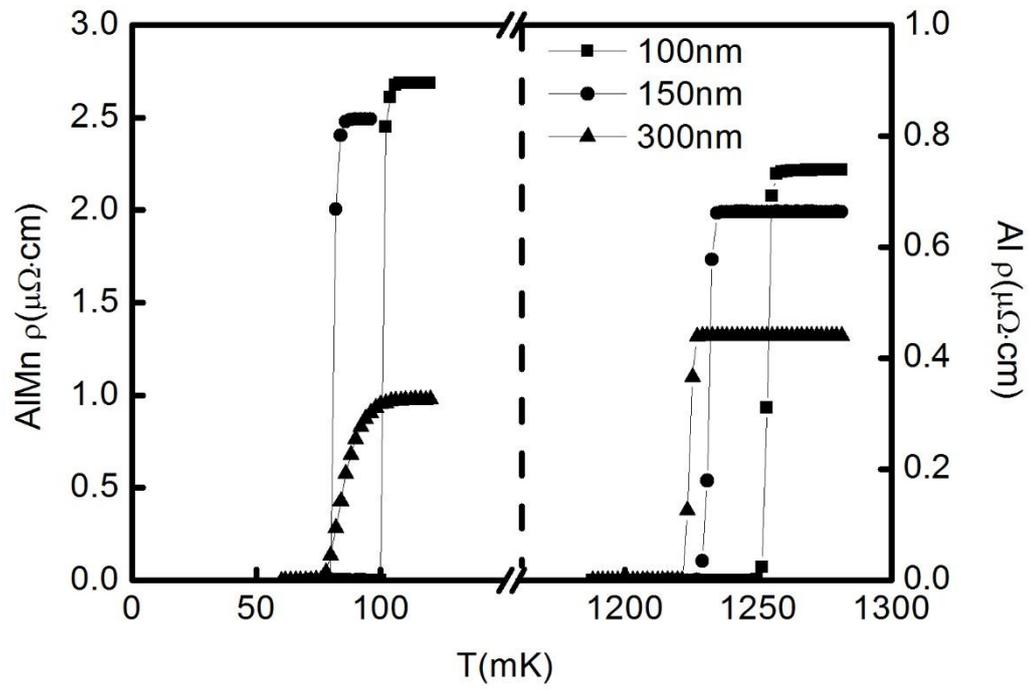

**Figure 2.** Resistivity versus temperature curves for undoped (right panel) and Mn-doped (left panel) aluminum films. Squares, circles and triangles represent 100-, 150-, and 300- nm thick films.

# Figure 3

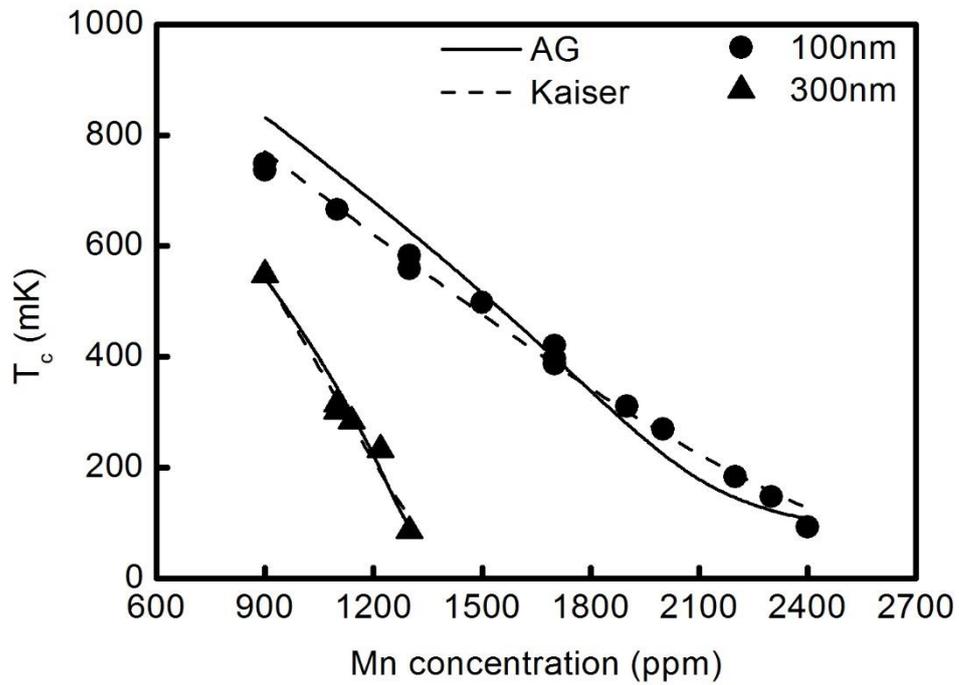

**Figure 3.** Superconducting transition temperature of ion-implanted AlMn films as a function of Mn doping concentration. Solid circles refer to 100-nm thick films and solid triangles refer to 300-nm thick films. Solid and dashed lines are fitting curves based on the AG and Kaiser model, respectively.

# Figure 4

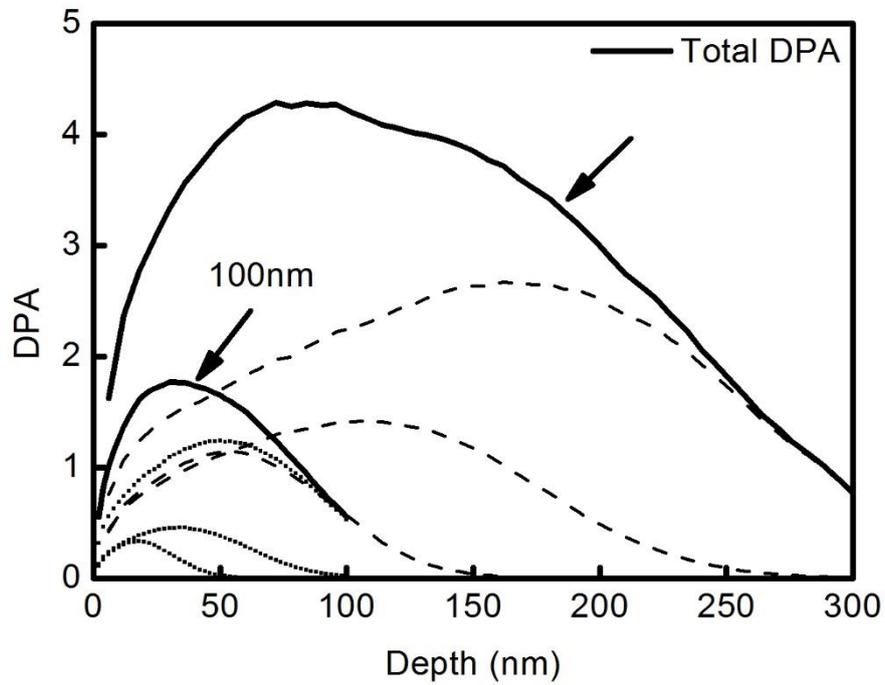

**Figure 4.** Simulated displacement per atom (DPA) as a function of film thickness for 100- and 300- nm thick AlMn films. Both films had a Mn doping concentration of 1100 ppm. Dashed curves represent simulation results for each individual implantation energy used in the implantation of 300-nm thick films; and dotted curves represent results for each energy used in the implantation of 100-nm thick films. Solid curves are the sum of dashed and dotted curves.

# Figure 5

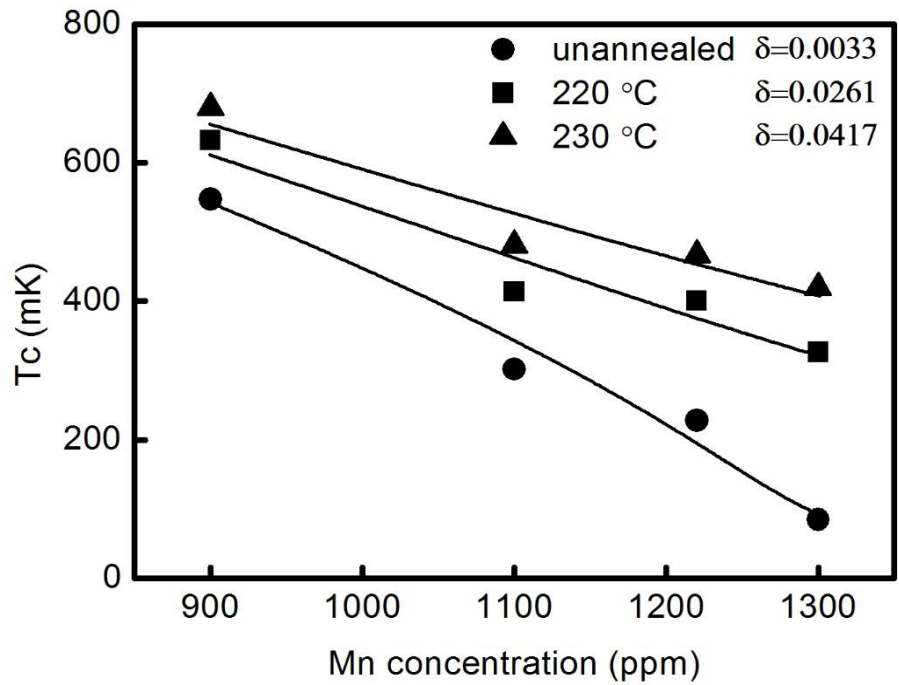

**Figure 5.** Superconducting transition temperature of 300-nm thick ion-implanted AlMn films as a function of Mn doping concentration. Solid circles refer to unannealed samples, solid squares refer to samples annealed at 220 °C and solid triangles refers to samples annealed at 230 °C. $\delta$ is the fitted exchange coupling parameter.